\documentclass[11pt]{article}
\usepackage{jr}
\def\bx{\mathbf{x}}
\def\bt{\mathbf{t}}
\def\bbeta{\boldsymbol{\beta}}
\def\bdel{\boldsymbol{\delta}}

\begin{document}

\title{Re-calibration of sample means.}
\author{E. Greenshtein\thanks{Israel Central  Bureau of Statistics. eitan.greenshtein@gmail.com}\;\; and Y. Ritov\thanks{Department of Statistics, The Hebrew University of Jerusalem. yaacov.ritov@gmail.com}}
\maketitle

\begin{abstract}

We consider the problem of calibration and the GREG method as suggested and studied in Deville and Sarndal (1992).
We show that a GREG type estimator is typically not minimal variance unbiased estimator even asymptotically. We suggest
a similar estimator which is unbiased but is asymptotically with a minimal variance.

\end{abstract}

\section{Introduction}

The purpose of this note is to examine the popular calibration techniques, suggested, e.g., in  Deville and Sarndal
(1992), or Sarndal et.al. (1992) Chapter 6.4, those calibrated estimators are also known as GREG (the general regression estimator). Our development and criterion are elementary. We are interested in finding
a minimum variance linear estimator.This leads lead to a very similar to the GREG estimator in form estimator, but with different constants. The difference between these two approaches as demonstrated in what follows. This demonstration is the main purpose of this note.

First we review the above mentioned calibration GREG approach, following  Deville and Sarndal (1992).
Consider a finite population $U=\{1,...,N \}$, and a sample $S$, $S \subset U$. Denote $\pi_i= P( S\ni i)$,
$\pi_{ij}=P( S \supseteq\{i,j\})$. Let $(y_i, \bf{x}_i)$, be quantities associtaed with item $i$, $i \in U$,
here $y_i$ is a scalar and $\bf{x}_i$ is a vector.  The quantity of interest is $t_Y= \sum_U y_i$, while
$\bf{x}_i$ are considered as covariates. Suppose the total ${\bf t}_X= \sum_U \bf{x}_i$ is known, w.l.o.g.,
${\bf t}_X=0$. Then, it is suggested to utilize that information about the totals through the following reasoning.

Define $\hat{\bf{t}}_X=\sum_{i \in S}  {\bf{x}_i}/{\pi_i}\equiv\sum_{i \in S} d_i \bf{x}_i$ and
$\hat{t}_Y= \sum_{i \in S} {y_i}/{\pi_i}\equiv \sum_{i \in S} d_i y_i$, where $d_i=1/\pi_i$.
The above are the  Horowitz-Thompson estimators, hence we have $E\hat{t}_y=t_Y$ and $E \hat{\bf{t}}_X={\bf t}_X=0$.
However, the value of $\hat{\bf{t}}_X$ is typically different than 0, which is unfortunate.

It is suggested to find ``better'' or ``improved'' weights $w_i, \; i \in S$ ("better" than $d_i$) and estimate $t_y$ by
$\sum_{i \in S} w_i y_i$. The heuristic derivation of the improved (random) weights $w_i, \; i \in S$
is the following. Given $S$ denote by $\bf{w}$ the vector of improved weights. Then $\bf{w}$ is defined as the solution of
the program:
\begin{equation}
\label{prog}
\begin{split}\min_{\boldsymbol{\omega}} \sum_{i \in S} (\omega_i - d_i)^2/d_iq_i \\
s.t. \sum _{i \in S} \omega_i \bx_i= 0;
\end{split}\end{equation}
here, the $q_i$ are selected parameters, which, as a default, suggested to be set to 1. The resulting estimator denoted $\hat{t}_{y|\mathbf x}$, may be written as: $\hat{t}_{y|\bx}= \sum w_k y_k$.

The solution of \eqref{prog} is simple. Using a vector of Lagrange multipliers $\lambda$ we can find that
\begin{align*}
\begin{split}
w_i&= (1+\lambda\t q_i\bx_i)d_i.
\end{split}
\end{align*}
where $\lambda$ is such that the constraint is satisfied, namely
\begin{align*}
\begin{split}
\lambda &= -\Bigl(\sum_{i\in S} q_id_i\bx_i\bx_i\t) \Bigr)^{-1} \sum_{i\in S}d_i \bx_i
\\
&\equiv - H_q^{-1}\hat t_X ,
\end{split}
\end{align*}
where $H_q=\sum_{i\in S} q_id_i\bx_i\bx_i$. Hence
\begin{align*}
\begin{split}
\hat t_{y|\bx} &= \hat t_Y -  \hat {\boldsymbol{\beta}}\t \hat t_X,
\end{split}
\end{align*}
where $\hat {\boldsymbol{\beta}}= H_q^{-1}\sum_{i\in S}d_iq_i y_i\bx_i$.

In the following we consider weights $q_i\equiv 1$, and denote then $H_q$ simply by $H$.

Note that for any (pre-determined) $\boldsymbol{\beta}$, $\hat t_Y-\boldsymbol{\beta}\t\hat t_X$ is an unbiased estimator of $t_Y$. Hence we may look for the minimal variance estimator of this type. One may restrict himself to a linear estimator (linear in $Y_i$, $i\in S$). That is, an estimator of the form $\sum w_i y_i$, with a sequence of weight that could simultaneously  be used for getting an estimator for any functional. Still one may look for such weights that would ensure that the estimator has a minimal variance. We will argue that the weights given by \eqref{prog} are generally speaking, far from being optimal.

Similar problem were discussed in Bickel, Klaassen, Ritov, and Wellenr (1998) in the context of \iid observations and semiparametric models. The question there, was defined as the semiparametric efficient estimation of parameter, when other parameters are known (e.g., estimation of the joint distribution, when the marginal distributions are known). Our solution is similar to the examples analyzed in that literature.



\section{Minimum variance linear unbiased estimator}

Consider estimators of $t_Y$ which are linear in $\hat{t}_Y$ and $\hat{t}_X$, i.e., of the form
$$ T(\boldsymbol{\beta})=\hat{t}_Y - \boldsymbol{\beta}\t \hat{\bt}_X,$$ where $\bbeta$ is non-random.
The above class is  unbiased since  $E\hat{\bt}_X=0$.
Consider the estimator $T(\bbeta_o)$ in the above class with minimal variance.
Clearly,
\begin{equation}\label{prog}
\boldsymbol{\beta}_o=\Sig_{\hat \bt_X}^{-1}\Sig_{\hat \bt_X,\hat t_Y},
 \end{equation}
where $\Sig_{\hat \bt_X}$ and $\Sig_{\hat \bt_X,\hat t_Y}$ are the variance-covariance matrix of $\hat \bt_X$, and the covariance vector of $\hat \bt_X$ and $\hat t_Y$, respectively.

First we argue that $\hat {\boldsymbol{\beta}}$ is not a consistent estimator of $\boldsymbol{\beta}_0$. The following example, while being extreme, is enlightening.

\begin{example}
 Consider a population divided into two stratas of equal sizes. For each $i \in U$ there is a corresponding $y_i$ and $x_i$, i.e., we have one dimensional covariates.
Suppose we randomly sample $n$ units from each strata, i.e., a total of $M=2n$ where $\pi_i \equiv {M}/{N}$.

Assume the mean of $x_i$ in stratum 1 is -1 and their mean in stratum 2 is 1.
The variance of $x_i$ within each stratum is $\sigma^2$. Now assume
in stratum 1, $y_i \equiv -1$, while in stratum 2, $y_i \equiv 1$.  Therefore, $\var(\hat t_Y)=0$, and hence $\boldsymbol{\beta}_o=0$. In fact, the optimal estimator in this case is simply $\hat t_Y\equiv 0$, on the other hand, $\hat\beta =H^{-1}\sum_{i\in S}d_i y_i\bx_i \cip (1+\sig^2)^{-1}$.  Asymptotically (as $n\to\en$ ) the GREG estimator $T(\hat\beta)\approx -(1+\sig^2)^{-1}\hat\bt_X$ has therefor variance of order $N^2/n$, while the optimal estimator for this case is exact with zero variance.
\end{example}

The difference between $\boldsymbol{\beta}_o$ and $\hat{ \boldsymbol{\beta}}$ would be large, when there is more than  a scale difference between the second moments of $\hat t_Y,\hat t_X$ and of those of $Y,X$. This precludes the simple random sample, but is typical for other sampling scheme.   The following example is less extreme than the first one, but describes a practical situation.

\begin{example} Suppose we sample clusters, the units in the sample are indexed by $j$ and $k$, where all units in cluster $j$, refer to the same central value $s_j$, and satisfy $x_{jk}=s_j+\eps_{jk}$ and
$y_{jk}=s_j+\gamma\nu_{jk}$, where the correlation between $\eps_{jk}$ and $\nu_{jk}$ is 0. Suppose that $K$ units are sampled in each cluster. It is clear that if the number of clusters is large, then with obvious notation: $\bbeta_o=\Sig_s/(\Sig_s+\Sig_\eps/K)$ while $\hat\bbeta \cip \Sig_s/(\Sig_s+\Sig_\eps)$. In the simple case where $\Sig_s=\Sig_\eps=\Sig_\nu$, if $K=5$ then the estimator with $\beta_0$ would have a variance smaller by approximately 25\% than the variance of the estimator using $\hat\beta$. The difference is approximately $50\%$ when $K=10$.

\end{example}

In order to estimate the  $\Sigma_{  \hat{ \bf{t}}_X}$ and $\Sigma_{\hat{t}_Y, {\bf  \hat{\bf{t}}_X}}$, we may use the classical
variance estimators for Horovitz-Thompson estimator, see,  e.g., Cochran (1977) or Sharon (1999). Those estimators are typically given in the literature for
one dimensional variance rather than to a covariance matrix, however the same reasoning applies.
 Since $\bt_X=0$,
\begin{align*}
\Sig_{\hat t_Y,{ \bf \hat t_X}} &= \E\sum_{i,j\in S} \frac1{\pi_i\pi_j}y_i\bx_j
\\
&=   \sum_{i,j\in U} \frac{\pi_{ij}}{\pi_i\pi_j}y_i\bx_j
\\
&=    \sum_{i,j\in U} \Bigl(\frac{\pi_{ij}}{\pi_i\pi_j}-c\Bigr)y_i\bx_j, \quad \forall c.
\end{align*}
Similarly,

\begin{align*}
\Sig_{\hat t_X}
&=    \sum_{i,j\in U} \Bigl(\frac{\pi_{ij}}{\pi_i\pi_j}-c\Bigr)\bx_i\bx_j\t, \quad \forall c.
\end{align*}
Hence, the following are unbiased estimators:
\begin{equation}
\label{unbiased}
\begin{split}
\hat \Sig_{\hat t_X,\hat t_Y}
&=    \sum_{i,j\in S} \frac1{\pi_{ij}}\Bigl(\frac{\pi_{ij}}{\pi_i\pi_j}-c\Bigr)y_i\bx_j, \quad \forall c.
\\
\hat\Sig_{\hat t_X}
&=    \sum_{i,j\in S} \frac1{\pi_{ij}}\Bigl(\frac{\pi_{ij}}{\pi_i\pi_j}-c\Bigr)\bx_i\bx_j\t, \quad \forall c.
\end{split}
\end{equation}

We assume that we consider a sequence of populations and designs such that the estimators in \eqref{unbiased} are consistent. Typically, taking $c=1$ in \eqref{unbiased}, would suffice to make most of the terms of lower order than the diagonal. In a simple random sample without replacement, $c=(n-1)N/n(N-1)$ leaves only the diagonal.

\begin{theorem}
Let $\hat{\boldsymbol{\beta}}_o = \hat\Sig_{\hat \bt_X}^{-1}\hat\Sig_{\hat t_Y,\hat \bt_X}$, where the terms in the RHS are given by \eqref{unbiased}  with a given $c$. Then
\begin{align*}
T(\hat\bbeta_o) &= \sum_{i\in S}w_i y_i,
\end{align*}
where
\begin{align*}
w_i&=  \frac{1}{\pi_i}-\hat \bt_X\t \hat\Sig_{\hat \bt_X}^{-1}\sum_{j\in S}\frac1{\pi_{ij}} \Bigl(\frac{\pi_{ij}}{\pi_i\pi_j}-c\Bigr)\bx_j,\quad i\in S.
\end{align*}
Thus the weights are a function of $\bx_i$, $\pi_i$, $\pi_{ij}$, $i,j\in S$ only. In particular $\sum_{i\in S}w_i \bx_i=0$.
\end{theorem}

\section{ Examining $\hat{ \boldsymbol{\beta}}$ under  linear model assumptions.}

In this section we will examine the rational in the estimator $\hat{\boldsymbol{\beta}}$ under the convenient and (too) often
assumed super-population model under which $Y_k = \boldsymbol{\beta}{\bf x}_k +\epsilon_k$, where $E \epsilon_k=0$
and for simplicity assume that $\epsilon_k$, $k \in U$ have equal variance.

Under this model it is easy to check that $\Sigma_{XY}= \boldsymbol{\beta} \Sigma_{XX}$, and
$\Sigma_{{\bf t}_X,t_Y}= \boldsymbol{\beta} \Sigma_{\bf{t}_X}$. Hence $\hat{\boldsymbol{\beta}}$ is a possible estimator of $\boldsymbol{\beta}_o=\boldsymbol{\beta}$. However, if this model is assumed, it is still not clear why $\hat{\boldsymbol{\beta}}$ should be used. We have here a  standard regression problem. Elementary regression theory (namely the  Gauss-Markov
Theorem) implies that the optimal estimator is not $\hat{\boldsymbol{\beta}}$, but the standard un-weighted linear regression of $Y_1,\dots,Y_n$ on $\bx_1,\dots,\bx_n$.

It might be argued that in fact we are taking the linear model super-population assumption with a grain of salt,
and thus we are using the estimator for
\begin{equation}
\boldsymbol{\beta}= \arg\min_{  {\bf b}} \sum_{i \in U} (y_i - {\bf b}\t{\bf x})^2, \label{eqn:nolinear}
\end{equation}
which is defined
under no linear model assumptions. However, since in this case we have no interest in that population parameter per se, but just in a tool for construction a good estimator for $t_Y$, than $\boldsymbol{\beta}_o$ should be our target.

To summarize. If we are interested in the super-population parameter, than $\hat{\boldsymbol{\beta}}$ is not efficient, and if we are interested in good estimator of $t_Y$, than $\hat{\boldsymbol{\beta}}$ is not consistent under complex sampling schemes.

\section{A partial knowledge of ${\bf t}_X$    }

In many cases ${\bf t}_X$ is not really known. However, it might be that there is an additional independent  sample
with information about ${\bf t}_X$ but not about $t_Y$. Thus we have three unbiased estimators $\hat t_Y^1,\hat { \bf t}_X^1$ based on one sample, and $\hat {\bf t}_X^2$ based on another independent sample. The best estimator of  $t_Y$ would be based on these three. Following the same argument as before we should consider estimator of the form
\begin{align*}
\hat \bt_Y^1 - \boldsymbol{\beta}\t(\hat {\bf t}_X^1 - \hat {\bf t}_X^2).
\end{align*}
Note that this estimator yields an unbiased estimate of $t_Y$ for any $\boldsymbol{\beta}$. The optimal value, however, is given by
\begin{align*}
\boldsymbol{\beta}_{o2}  &= -\bigl(\Sig_{\hat\bt^1_X} + \Sig_{ \bt^2_X}\bigr)^{-1}\Sig_{ {\hat\bt}^1_X,\hat\bt^1_Y}.
\end{align*}
Note, that if $\hat {\bf t}_X^2$ is based on the all universe $U$, then $\Sig_{\hat {\bt}^2_X}=0$, and $\beta_{o2}=\beta_o$.

Even more generally, we can consider a situation in which $\bx$ is measured for all units in the a super sample $S_2$, $ S_1\subseteq S_2\subseteq U$, while the $y$ values are measured only for units in the smaller sample $S_1$. For example, $y$ is measured only for one unit in a cluster, while the $\bx$ is measured for all units. Let $\hat\del_X=\bt_X^1-\bt_X^2$. It may be natural to assume that $\hat \del_X$ is correlated with $\hat t_Y$ while having a mean 0. We consider the natural extension $\hat t_Y^1 - \boldsymbol{\beta}_{o3}\t\hat\bdel_X$, with $\boldsymbol{\beta}_{o3}=-\Sig_{\hat\bdel_X}^{-1}\Sig_{\hat\bdel_X,\hat t_Y^1}$.
\begin{example}
Consider the super-population model in which it is assumed that   $y_{j,k}=\beta x_{j,k}+\eps_{j,k}$, $j=1,\dots,M$, $k=1,\dots,K$  where $\eps_{j,k}$ are \iid, independent of $x_{j,k}$, while $x_{j',k'}$ and $x_{j,k} $ are independent if $j\ne j'$, and have correlation $\rho$ if $j=j'$ and $k\ne k'$. Let $\var (x_{j,k})=\sig^2$.  Consider the sample $C\subset\{1,\dots,M\}$ of $n$ clusters.  Suppose that for each $j\in C$,  $x_{j,k}$, $k=1,\dots,K$ are obtained, while only $y_{j,1}$ is measured, assume also that   $M\gg n$. The universe size is $N=MK$. Hence, we assume for simplicity a simple random sample (with replacement) of clusters.
Then
\begin{align*}
\hat \del_X &= \frac {N}n \sum_{j\in C}\bigl( x_{j,1}-K^{-1}\sum_{k=1}^K x_{j,k}  \bigr)
\end{align*}

It is easy to verify
\begin{align*}
\var(\hat t_Y^1) &= \frac{N^2}n \beta^2\sig^2
\\
\var(\hat \del_X) &= \frac{N^2}n  \frac{K-1}K (1-\rho)\sig^2
\\
\cov(\hat \del_X,\hat t_Y^1) &= \frac{N^2}n    \frac{K-1}K (1-\rho)\beta\sig^2.
\end{align*}
Hence
\begin{align*}
\frac{\var(\hat t_Y^2-\beta_{o3}\hat\del_X)}{\var(\hat t_Y^2)} &= 1-\frac{K-1}K(1-\rho).
\end{align*}
The efficiency of the scheme increases as $K$ increases and $\rho$ decreases. Note that the case of a simple random sample of units in which the $y$ value is measured only for a small random sub-sample, corresponds to $\rho=0$.
\end{example}

\bigskip

{\bf References}

\begin{list}{}{\setlength{\itemindent}{-1em}}

\item
Bickel, P.J., Klassen, C.A.J., Ritov, Y., Wellner, J.A. (1998). \emph{Efficient
and adaptive estimation  for semiparametric models}. Springer Verlag New York Inc.

\item
Cochran, W.G (1977). \emph{Sampling techniques}. 3d ed. New York: Wiley.

\item
Deville, J.C. and Sarndal, C.E. (1992). Calibration estimators in survey sampling. {\it JASA}, {\bf 87}, 376-382.

\item
Sahron L. Lohr (1999). \emph{Sampling Design and Analysis}. Brooks/Cole publishing company.

\item

Sarndal, C. E., Swenson, B., Wretman, J. (1992). \emph{Model assisted survey sampling}. New York: Springer-Verlag.

\end{list}

\end{document}